\documentclass[twocolumn]{aastex62}
\usepackage{amssymb}
\usepackage{amsmath}
\usepackage{array,multirow}
\usepackage{comment}
\usepackage{enumerate}
\usepackage{bm}

\newcommand{\Msun}{\ifmmode {M_{\odot}}\else${M_{\odot}}$\fi}
\newcommand{\Rsun}{\ifmmode {R_{\odot}}\else${R_{\odot}}$\fi}
\newcommand{\Porb}{\ifmmode {P_{\rm orb}}\else${P_{\rm orb}}$\fi}
\newcommand{\dart}{\texttt{dart\_board}}
\newcommand{\cosmic}{{\texttt{COSMIC}}}
\newcommand{\emcee}{{\texttt{emcee}}}
\newcommand{\bigG}{\ifmmode {\mathcal{G}}\else${\mathcal{G}}$\fi}

\newcommand{\SUPA}{\affiliation{SUPA, School of Physics and Astronomy, University of Glasgow, Glasgow G12 8QQ, UK}}
\newcommand{\CIERA}{\affiliation{Center for Interdisciplinary Exploration and Research in Astrophysics (CIERA), 1800 Sherman Ave., Evanston, IL, 60201, USA}}
\newcommand{\NU}{\affiliation{Department of Physics and Astronomy, Northwestern University, 2145 Sheridan Rd., Evanston, IL 60208, USA}}

\submitjournal{AAS Journals}

\shorttitle{Modeling GW150914's source with \dart}
\shortauthors{Andrews et al.}

\begin{document}

\title{{\bf Targeted modeling of GW150914's binary black hole source with \dart}}

\author[0000-0001-5261-3923]{Jeff J.\ Andrews}
\CIERA
\email{jeffrey.andrews@northwestern.edu}

\author[0000-0003-1172-5755]{Julianne Cronin}
\NU

\author[0000-0001-9236-5469]{Vicky Kalogera}
\CIERA
\NU

\author[0000-0003-3870-7215]{Christopher P.\ L.\ Berry}
\CIERA
\NU
\SUPA

\author[0000-0001-8952-676X]{Andreas Zezas}
\affiliation{Foundation for Research and Technology -- Hellas, IESL, Voutes, 71110 Heraklion, Greece}
\affiliation{Physics Department \& Institute of Theoretical \& Computational Physics, University of Crete, 71003 Heraklion, Crete, Greece}
\affiliation{Center for Astrophysics, Cambridge, MA, USA}

\begin{abstract}
We present a new method to extract statistical constraints on the progenitor properties and formation channels of individual gravitational-wave sources. Although many different models have been proposed to explain the binary black holes detected by the LIGO Scientific and Virgo Collaboration (LVC), formation through isolated binary evolution remains the best explored channel. Under the assumption of formation through binary evolution, we use the statistical wrapper \dart\ coupled with the rapid binary evolution code \cosmic\ to model the progenitor of GW150914, the first gravitational-wave signal detected by the LVC. Our Bayesian method combines the likelihood generated from the gravitational-wave signal with a prior describing the population of stellar binaries, and the Universe's star-formation and metallicity evolution. We find that the dominant evolutionary channel for GW150914 did not involve a common-envelope phase, but instead the system most probably ($\simeq$70\%--90\%) formed through stable mass transfer. This result is robust against variations of various model parameters, and it is reversed only when dynamical instability in binaries becomes more likely when a strict condition favoring common envelopes is adopted. Our analysis additionally provides a quantitative description of the progenitors relevant to each channel. 
\end{abstract}


\section{Introduction}
\label{sec:intro}

One of the breakthroughs of twenty-first century physics has been the observation by the Laser Interferometer Gravitational-Wave Observatory (LIGO) of gravitational waves caused by the coalescence of two black holes \citep[BHs;][]{GW150914}.
Since then the LIGO \citep{aLIGO} and Virgo \citep{AdV} observatories have detected many others, such that there are now $\simeq40$ observations of the coalescence of two BHs \citep{GWTC-2}. Additionally, there are two observations from the coalescence of two neutron stars \citep[NSs;][]{GW170817, GW190425} and the possible mergers of a BH and a NS \citep[GW190814, GW190426\textunderscore152155;][]{GW190814, GWTC-2}. Gravitational-wave sources observed by the LIGO Scientific and Virgo Collaboration (LVC) have ushered in a new chapter of physics, including constraints on the equation of state of NSs  \citep[][]{GWTC-1,malik2018,coughlin2019,gw170817-model-selection,landry2020}, tests of general relativity \citep{gw150914-TGR,yunes2016,LIGO_GR_tests}, and the production of heavy elements \citep{gw170817-MMA,chornock2017,tanvir2017,rosswog2018,watson2019}.

In addition to their many contributions to fundamental physics, these observations also afford unique astrophysical insights, providing information related to compact object formation. Several formation channels exist, including the merger of primordial BHs \citep{bird2016, sasaki2016,clesse2020}, the homogeneous evolution of two massive stars through tidal locking \citep{mandel2016, de_mink2016, marchant2016} or mass transfer \citep{eldridge2016}, the evolution of two massive, population III stars \citep{kinugawa2014, kinugawa2021}, a dynamical origin in which exchange interactions produce BHs in tight orbits at the centers of dense stellar environments such as globular clusters \citep{rodriguez2016a, hong2018, rodriguez2019, martinez2020, antonini2020}, young stellar clusters \citep{ziosi2014, di_carlo2019, santoliquido2020} and nuclear star clusters \citep{antonini2016b, stone2017, secunda2019,grobner2020,fragione2020} as well as the interaction of triple (or higher order) systems \citep{antonini2017a, rodriguez2018, antonini2018, fragione2019}. However, perhaps the most developed channel involves the formation of BHs from two high-mass stars evolving together in a binary system \citep{belczynski2016, stevenson2017, eldridge2016, spera2019, bavera2020}; in this channel the stars undergo multiple phases of mass transfer (MT), tidal coupling, and survival through two core-collapse events.\footnote{While the chemically homogeneous channel \citep[e.g.,][]{de_mink2016} also involves the evolution of two massive stars in a binary system, here we are referring to more traditional binary evolution channels in which the two stars' orbit is not so tight that strong internal mixing is induced.} 

No systems comprised of two BHs are known to exist in our Galaxy, so the LVC detections form a unique dataset. The few known high mass X-ray binaries might be expected to evolve into binary BHs, but these will likely form into systems too wide to merge in a Hubble time, and therefore do not form progenitors of LVC detections \citep{qin2019,neijssel2021}. Although there have been claims that several Wolf--Rayet X-ray binaries are the progenitors to binary BH mergers \citep{bulik2011, belczynski2013, bogomazov2014}, the optically thick winds in these systems make the nature of the compact-object accretor uncertain. Recently, \citet{mondal2020} argued that a substantial fraction of the ultra-luminous X-ray binaries observed throughout the Universe will evolve into merging BH binaries; however, these systems are distant, and, despite their growing population \citep{kovlakas2020}, ultra-luminous X-ray binaries are still observationally ill-constrained. Therefore, we are largely informed about BH--BH formation using model predictions. By comparing the observed sample with a synthetic population generated using rapid binary evolution codes, the BH masses and spins in the LVC sample have been used to constrain binary evolution \citep{eldridge2016, belczynski2016b, stevenson2017, giacobbo2018a, giacobbo2018b, qin2018, Neijssel2019, spera2019, stevenson2019, belczynski2020, zevin2020, bavera2020a, bavera2020, bouffanais2020, santoliquido2020, wong2020, zevin2020b}.

Gravitational-wave observations have been used to both understand the properties of their source progenitors and to constrain the physics that governs compact-object mergers and the production of gravitational radiation. In this work, we will address the first question: What were the progenitor stars that evolved to form a merging BH--BH binary? Previous investigations into this question have derived results from a subset of a synthetic population selected such that the chirp mass (perhaps with other parameters included as well) falls within some range of the LVC-derived values. Such a technique has been employed to provide models for the formation of GW150914 \citep{belczynski2016} as well as several other systems \citep[e.g.,][]{stevenson2017, GW170817_formation, olejak2020, zevin2020}. While this method can provide approximate constraints on the formation of individual systems, it does not make full use of the information available from the gravitational-wave observation. First, this approach does not take into account the likelihood of different source parameters, and all BH masses within the range of uncertainties are treated as equal. Second, the parameter values provided by the LVC are inferred using uninformative prior probabilities for the source properties which may be a poor reflection of the underlying astrophysical distribution. The inferred values can be sensitive to the choice of prior \citep[e.g.,][]{GWTC-1, fishbach2020, zevin2020a}, and so to be self-consistent, one should use a prior which matches the population model being considered. In \citet{andrews2018}, we described a code \dart\ that uses a Bayesian method to model the formation of individual stellar binaries and produce posterior probability distributions of the set of progenitor properties, and here we apply this approach to find the probable source for GW150914.

In Section~\ref{sec:method} we describe how we adapt \dart\ to model individual gravitational-wave sources using the parameter-estimation posterior samples provided in the publicly available LVC data \citep{GWTC-1}.\footnote{Posterior samples for GW150914 available from \href{https://doi.org/10.7935/KSX7-QQ51}{doi.org/10.7935/KSX7-QQ51}.} We describe the basic formation channel we find for GW150914 in Section~\ref{sec:results}, and describe its robustness when accounting for the principle uncertainties in binary evolution physics.\footnote{For reproducibility, we make publicly available the scripts we used for running our separate models as well as those used in creating the figures displayed throughout this work: \href{https://github.com/astroJeff/dart_GW150914}{github.com/astroJeff/dart\_GW150914}} We comment on the relevance of our results and their implications for current and future gravitational-wave sources in Section~\ref{sec:conclusion}.

\section{Method}
\label{sec:method}

\subsection{Model Outline}

To model individual gravitational-wave detections, we start with the rapid binary evolution code \cosmic\ \citep{breivik2020}. This code is an open-source and updated form of the rapid binary evolution code {\tt BSE} \citep{hurley2002}, which itself uses fitting formulas to model the evolution of single stars \citep{hurley2000}. \cosmic\ models the evolution of binary stars using the best possible physical prescriptions, starting from the zero-age main sequence, through tidal interactions, multiple MT phases, and supernovae until the stars in these systems have completely evolved into compact objects. \citet{breivik2020} provides a comprehensive description of the updated physics included within \cosmic. Here, we only briefly mention the relevant, significant choices in physics we have made for our default model. During stable MT, we assume an accretor accepts half the mass lost by the donor (we discuss this assumption below in Section \ref{sec:model_variations}). For systems with donor masses more than three times the accretor, the binary will enter a common envelope (CE). Despite their not-yet-fully-developed core, Hertzsprung gap stars are allowed to proceed through a CE, although we vary this assumption below. Evolution follows the $\alpha$--$\lambda$ prescription when going through a CE, with $\alpha = 1$ \citep{webbink1983, de_marco2011} and $\lambda$ set by the prescription in \citet{claeys2014}. For sufficiently massive He-stars at core collapse, BHs are formed according to the Rapid prescription from \citet{fryer2012}, where upon core collapse we have allowed for up to $10\%$ of the baryonic mass of the core to be lost due to neutrinos, with the prescription described in Appendix~A.2 of \citet{zevin2020}. We adopt a prescription for pulsational pair-instability supernova that only affects stars forming BHs $\gtrsim$45 \Msun, making this process irrelevant for GW150914 in our model. At birth, these BHs are assumed to have small, but non-zero, natal kicks distributed isotropically, with a magnitude following a Maxwellian distribution with a dispersion velocity $\sigma_k = 10~\mathrm{km\,s^{-1}}$, and BH kicks are not modulated by fallback. Although the preceding options define the physics in our standard model, \cosmic\ is written to flexibly allow for variations in many of these prescriptions, several of which we explore in Section~\ref{sec:model_variations}.

Typically, rapid binary evolution codes like \cosmic\ are used to generate a large number ($>10^6$) of samples, from which a small subset is selected that produces binaries consistent with a class of objects or a particular observation. Rather than generate a synthetic population, we use \cosmic\ to model binaries one-at-a-time with initial binaries determined by our statistical wrapper \dart\ \citep{andrews2018}. This code considers binary evolution as a parameter-estimation problem, solving for the most probable zero-age main sequence (ZAMS) component masses, orbital period, eccentricity, and metallicity. To satisfy detailed balance, the SN kick velocity, direction, and mean anomaly for each star are included as model parameters.\footnote{At the moment of supernova, most but not all binaries have circularized prior to core collapse.} The result is that there is no randomness associated with binary evolution, and a particular combination of model parameters will evolve into the exact same binary every time it is run. For the binary BH sources we are interested in, our model contains $13$ parameters, which we summarize as $x_\mathrm{i}$.

\dart\ uses \emcee\ \citep{emcee}, a Markov-chain Monte Carlo (MCMC) implementation of the ensemble algorithm developed by \citet{goodman2010} to sample the parameter space based on a posterior probability, $P(x_\mathrm{i} | D, \Lambda)$ where $D$ refers to some set of observational data (in our case, the gravitational-wave signal) and $\Lambda$ refers to the parameters of the binary evolution model. The posterior probability can be split into a prior probability $P(x_\mathrm{i} | \Lambda)$ and likelihood $P(D | x_\mathrm{i}, \Lambda)$ using Bayes' Theorem:
\begin{equation}
    P(x_\mathrm{i} | D, \Lambda) = \frac{P(D | x_\mathrm{i}, \Lambda) P(x_\mathrm{i} | \Lambda)}{P(D | \Lambda)}.
\end{equation}
The denominator $P(D | \Lambda)$ represents the Bayesian evidence, which for the purposes of this study can be ignored since we are not comparing between models. Therefore, this method requires prior distributions on all the model parameters, as well as a likelihood function for gravitational-wave observations \citep{cutler1994}. 
Our analysis additionally needs to account for the evolution of the star-formation rate and metallicity of the Universe over cosmological times. 
We separately describe how we account for each of these terms below.

\subsection{Prior Function}
\label{sec:prior}

For each of our model parameters, we must provide a prior probability. We use standard priors, the primary star's mass $M_1$ is drawn from a Salpeter initial mass function \citep[IMF;][]{salpeter1955}, the secondary star's mass $M_2$ is drawn from a flat mass ratio, the orbital separation $a$ is drawn from an Opik's law distribution \citep{opik1924}, and we use a thermal eccentricity distribution:
\begin{equation}
\begin{array}{rlr}
    P(M_1) &\propto M_1^{-2.35}, & 8 < M_1/\Msun < 150, \\
    P(M_2 | M_1) &\propto M_1^{-1}, & 2 < M_2/\Msun < M_1/\Msun, \\
    P(a) &\propto a^{-1}, & 10 < a/\Rsun < 10^5, \\
    P(e) &\propto e, & 0 < e < 1.
\end{array}
\label{eq:priors_1}
\end{equation}
We find that these limits are sufficient to fully encompass the regime of systems producing binary BH mergers.

BHs receive SN kicks at birth with a magnitude following a Maxwellian distribution isotropically directed with a polar angle $\theta$, an azimuthal angle $\phi$, and a mean anomaly at the moment of collapse $\omega$:
\begin{equation}
\begin{array}{rlr}
P(v_k) &\propto v^2_k / \sigma^3_k \exp [-v_k^2/2\sigma_k^2], & 0<\sigma_k<\infty, \\
P(\theta) &\propto \sin \theta, & 0<\theta<\pi, \\
P(\phi) &\propto 1, & 0 < \phi < \pi, \\
P(\omega) &\propto 1, & 0 < \omega < 2 \pi.
\end{array}
\label{eq:priors_2}
\end{equation}
Since our algorithm relies only on the relative probabilities, improper priors are sufficient for our purposes, and for concision we neglect normalizing constants in each of these prior probabilities in Eq.~\eqref{eq:priors_1} and Eq.\eqref{eq:priors_2}. 

Our final model parameter, the metallicity $Z$, is determined in a different way, as it accounts for the cosmological evolution of the star-forming activity of the Universe. We discuss how we calculated that term in Section~\ref{sec:cosmology}.

\subsection{Including Cosmological History}
\label{sec:cosmology}

Both the average star formation rate and the typical metallicity of newborn stars within the Universe evolve over cosmic time. We use the star-formation rate as a function of redshift ($z$) from \citet{madau2014}:
\begin{equation}
    \psi(z) = 0.015 \frac{(1+z)^{2.7}}{1 + [(1+z)/2.9]^{5.6}}\ \Msun\ {\rm yr}^{-1}\  {\rm Mpc}^{-3}.
\end{equation}
Since this is a volumetric star-formation rate, we need to account for the increasing volume probed as a function of comoving distance $D_{\rm C}$, by multiplying the star formation rate by the comoving volume element $4 \pi D_{\rm C}^2\ {\rm d}D_{\rm C}$. The probability of a binary forming at a distance $D_{\rm C}$ is proportional to this volume element and the star formation rate at the redshift corresponding to that volume element \citep[in a flat space-time, the line-of-sight co-moving distance is equal to the transverse co-moving distance;][]{hogg1999}:
\begin{equation}
    P(D_{\rm C}) \propto 4 \pi D_{\rm C}^2 \frac{\psi(z)}{1+z},
    \label{eq:P_D_C}
\end{equation}
and the $(1+z)^{-1}$ term in Eq.~\eqref{eq:P_D_C} translates the volumetric star-formation rate from a local rate to an observer-frame rate. 

We also calculate the metallicity evolution of the Universe $Z(z)$:
\begin{equation}
    Z(z) = y \frac{\rho_*(z)}{\rho_b},
\end{equation}
where $y=0.23$ is a constant determined by integrating the metallicity yields of massive stars over a Salpeter IMF from $10 \Msun$ to $60 \Msun$, $\rho_*(z)$ is the mass density of stars at a redshift $z$, and $\rho_b=2.77 \times 10^{11} \Omega_b h^2 \Msun~\mathrm{Mpc}^{-3}$ is the baryon density of the Universe \citep{madau2014}. Using results from the {\it Planck} mission, we set $\Omega_b h^2=0.0223$ \citep{planck2016}. Finally, $\rho_*(z)$ is calculated by integrating the star formation rate over cosmic history from a very high redshift (we choose $z=200$ as our upper limit, but the exact value has a negligible impact) to $z$:\, 
\begin{equation}
\rho_*(z) = (1-R) \int_z^{200} \psi(z') \frac{{\rm d}z'}{H(z') (1+z')},
\end{equation}
where $H(z)$ is the Hubble parameter and $R$ is the average fraction of mass from a star that is returned to the interstellar medium, and $H$ is the Hubble parameter. Adopting the same assumptions on the IMF as in calculating $y$ above, we set $R=0.29$ \citep{madau2014}.

The prior on $Z$ is then determined to be a truncated normal distribution in log-space, around $Z(z)$, bound between $Z_{\rm min}$ and $Z_{\rm max}$, with a standard deviation of half a decade. 
\begin{equation}
    P(Z | z) = \begin{cases}    \mathcal{N}(\log Z; \log Z(z), 0.5), & Z_{\rm min} < Z < Z_{\rm max}, \\
    -\infty, & \text{other}.
    \end{cases}
\end{equation}
We set $Z_{\rm min} = 5\times10^{-5}$ and $Z_{\rm max} = 3\times10^{-2}$ based on the limits of applicability of the stellar evolution models within \cosmic. Throughout this work, $Z$ is considered to be in absolute terms ($X+Y+Z=1$) rather than relative to the solar metallicity.

We include these two functions as priors on our model parameters, $D_{\rm C}$ and $Z$: $P(D_{\rm C},Z) = P(Z | D_{\rm C})P(D_{\rm C})$.

\subsection{Including gravitational-wave observations}
\label{sec:LIGO}

\begin{figure}
\begin{center}
\includegraphics[width=\columnwidth,angle=0]{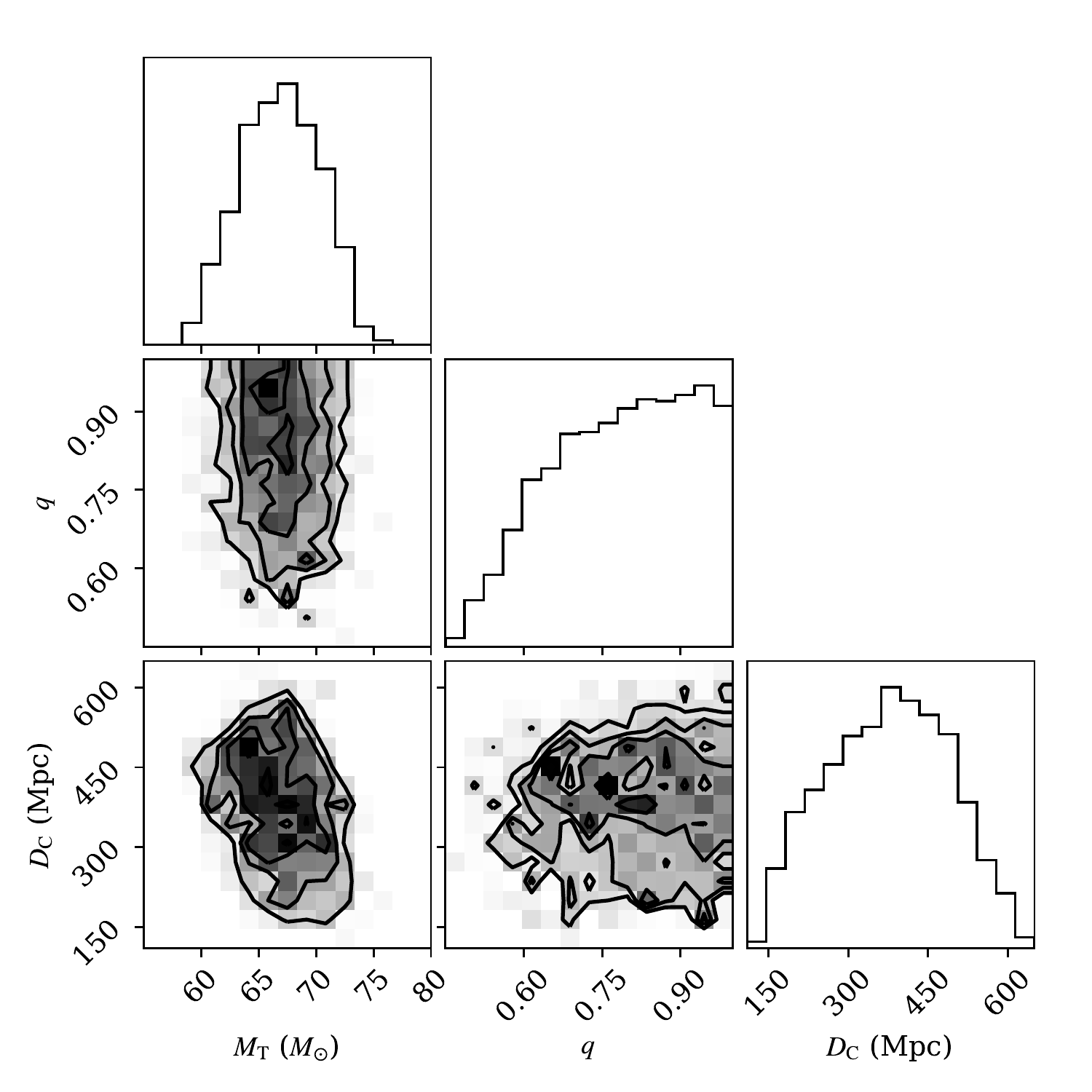}
\caption{ The distribution of publicly available samples provided by the LVC for GW150914. These have been weighted by the inverse of the astrophysical priors adopted by the LVC team and reparameterized, so this distribution corresponds to the likelihood of total mass, mass ratio, and comoving distance $P(M_{\rm T}, q, D_{\rm C})$. There is a strict boundary $q=1$, requiring mass ratios less than unity. \label{fig:1}}
\end{center}
\end{figure}

For each source detected, the LVC provides posterior samples on the two BH masses, as well as the luminosity distance to the merger, produced from their parameter-estimation pipeline \citep{gw150914-PE}. 
Here we do not incorporate spin constraints \citep[e.g.,][]{zevin2020b}. We reparameterize the LVC samples from individual BH masses and luminosity distance into a total mass $M_{\rm T}$, a mass ratio $q$ ($0\leq q \leq 1$), and a co-moving distance $D_{\rm C}$. These samples cannot be used directly in our calculations, as they were produced under the assumption of a specific prior distribution. The LVC provides a set of prior samples in this three-dimensional space, from which we derive a kernel density estimate (KDE) representation.\footnote{We use the code {\tt kalepy} \citep{kelley2021} to calculate the prior probability density, rescaling the total mass and distance so they range between $0$ and $1$.}
For each posterior sample provided by the LVC, we can calculate its associated prior probability using our KDE representation. By weighting each $j$th posterior sample by the inverse of the prior probability it was produced with,
\begin{equation}
    w_j = P(M_{{\rm T},j}, q_j, D_{{\rm C},j})^{-1}, \label{eq:weights}
\end{equation}
the set of posterior samples then represent the likelihood function. Figure \ref{fig:1} shows the joint likelihood function on $M_{\rm T}$, $q$, and $D_{\rm C}$ derived from these samples that we use throughout this work.

Using the binary evolution code \cosmic, we calculate the evolution of a set of initial binary parameters at ZAMS $x_\mathrm{i}$ until either the system produces two compact objects, merges, or reaches a Hubble time, whichever happens first. We summarize the final state of the binary as $x_\mathrm{f} = f(x_\mathrm{i})$, where $f$ is a abstract mapping representing the evolution of the binary from its initial state to the final state.

We start by marginalizing over the possible values of $M_{\rm T}$, $q$, and $D_{\rm C}$:
\begin{eqnarray}
    P(D | x_\mathrm{i}, \Lambda) &=& \int {\rm d}\Omega P(D, M_{\rm T}, q, D_{\rm C} | x_\mathrm{i}) \nonumber \\
     &=& \int {\rm d}\Omega P(D | M_{\rm T}, q, D_{\rm C}) P(M_{\rm T}, q, D_{\rm C} | x_\mathrm{i}), \nonumber \\
     & &
    \label{eq:likelihood_integral}
\end{eqnarray}
where ${\rm d}\Omega = {\rm d}M_{\rm T}\ {\rm d}q\ {\rm d}D_{\rm C}$.

Using the LVC samples drawn from the posterior probability $P(M_{\rm T}, q, D_{\rm C} | D)$, we can approximate the first term of the integrand in Eq.~\eqref{eq:likelihood_integral} by applying the weights $w_i$ defined in Eq.~\eqref{eq:weights}:
\begin{equation}
   P(D | M_{\rm T}, q, D_{\rm C}) \propto w\ P(M_{\rm T}, q, D_{\rm C} | D).
\end{equation}
Therefore, we can approximate Eq.~\eqref{eq:likelihood_integral} with a sum:
\begin{equation}
    P(D | x_\mathrm{i}, M) \approx \frac{1}{N} \sum^N_{j=0} w_j\ P(M_{T,j}, q_j, D_{{\rm C}, j} | x_\mathrm{i}), 
    \label{eq:likelihood_sum}
\end{equation}
where the subscript $j$ denotes samples drawn from the LVC posterior set $(M_{\rm T}, q, D_{\rm C})$, and $w_j$ are the prior weights on each of those samples. For each binary with initial conditions $x_\mathrm{i}$ and each LVC sample $\{M_{T,j}, q_j, D_{{\rm C},j}\}$, we calculate $P(M_{\rm T}, q, D_{\rm C} | x_\mathrm{i})$ by separating it into three terms:
\begin{equation}
    P(M_{T,j} q_j, D_{{\rm C},j} | x_\mathrm{i}) = P(M_{T,j} | M_{\rm T}) P(q_j | q)\ P(D_{{\rm C}, j} | x_\mathrm{i}),
    \label{eq:p-final-initial}
\end{equation}
where $M_{\rm T}$ and $q$ are the total system mass and mass ratio, respectively, of the final system: $(M_{\rm T}, q) \in x_\mathrm{f}$. This separation is justified since any covariance between $M_{\rm T}$, $q$, and $D_{\rm C}$ is naturally included within the posterior samples and associated weights. 

To calculate the first two terms on the right-hand side of Eq.~\eqref{eq:p-final-initial}, we use Gaussian distributions, which is mathematically equivalent to representing the posterior samples from the LVC as a KDE. We use a bandwidth of $1/100$ for $q$ and $1/10 \Msun$ for $M_{\rm T}$, and we add a reflective boundary to the distribution over $q$ at $q=1$.

The third term in the right-hand side of Eq.~\eqref{eq:p-final-initial} deals with the co-moving distance of the LVC observation. The total age of the progenitor is the sum of the system's formation time (time between ZAMS and when the second BH is formed), the merger time of the system due to gravitational radiation after the second BH is formed, and finally the lookback time associated with the system's comoving distance. 
Therefore, the probability of the total age of the system is set by the star formation rate of the Universe when the system formed.

We transform the probability from the co-moving distance to the merger to $D_{\rm C}^*$, the co-moving distance to the formation of the system (corresponding to the total age $t_{\rm formation} + t_{\rm merge} + t_{\rm lookback}$, where $t_{\rm lookback}$ is the lookback time to the merger): 
\begin{equation}
P(D_{{\rm C}, j} | x_\mathrm{i}) = P(D_{\rm C}^* | Z). \label{eq:transformation}
\end{equation}
Typically $t_{\rm formation}$ is significantly less than both $t_{\rm lookback}$ and $t_{\rm merge}$, but we include it here for completeness. The transformation between variables in Eq.~\eqref{eq:transformation} is allowed since the associated Jacobian is unity. The dependence on $Z$ comes from the joint evolution of the star formation rate and metallicity of the Universe over cosmological times. 

We have purposefully neglected to provide a prior probability for the metallicity $P(Z)$ in Section~\ref{sec:prior}, but this term is necessary, as $Z$ is a model parameter. 
A prior for the metallicity can be derived from the formation redshift of the source:
\begin{equation}
    P(Z, D_{\rm C}^*) =  P(Z | D_{\rm C}^*)P(D_{\rm C}^*),
\end{equation}
where both terms on the right-hand side are provided in Section~\ref{sec:cosmology}.

By combining the prior probabilities in Eq.~\eqref{eq:priors_1} and Eq.~\eqref{eq:priors_2} with the likelihood in Eq.~\eqref{eq:likelihood_sum}, we can now calculate the posterior probability for any $x_\mathrm{i}$ for a set of posterior samples provided by the LVC.

\begin{figure}
\begin{center}
\includegraphics[width=\columnwidth,angle=0]{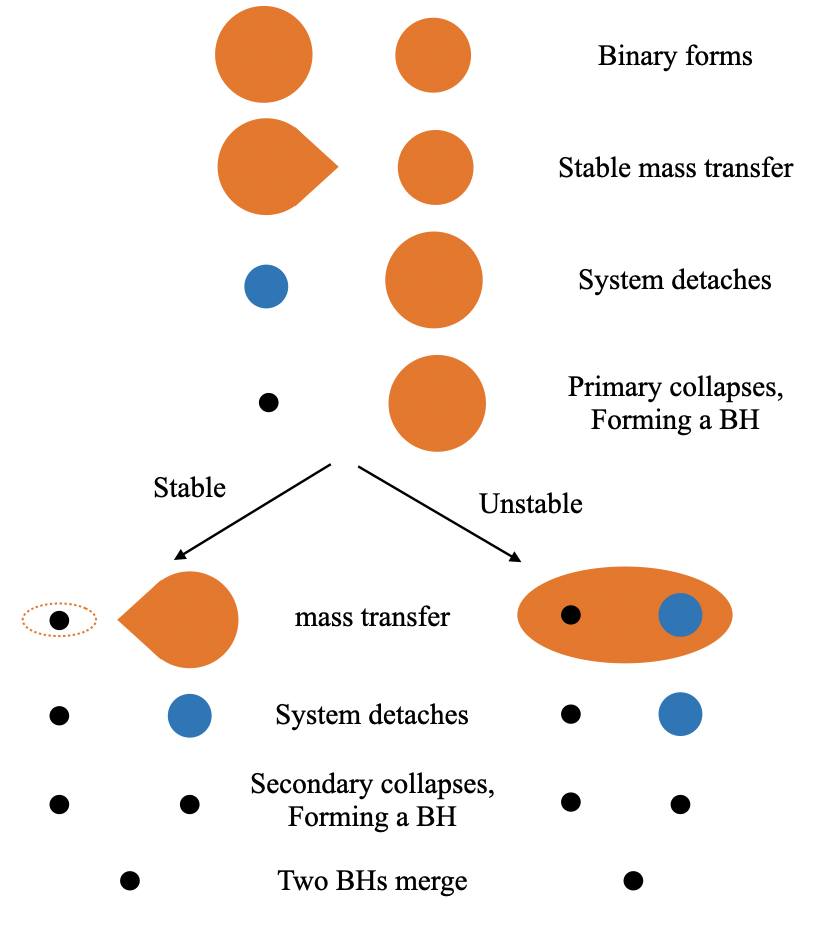}
\caption{ The two standard formation channels for forming binary BH mergers. In the CE channel, MT is unstable which requires orbits that are initially much wider than those of the stable MT channel. The divergence between the two channels depends on assumptions about MT stability. \label{fig:2}}
\end{center}
\end{figure}

\begin{figure*}
\begin{center}
\includegraphics[width=0.9\textwidth,angle=0]{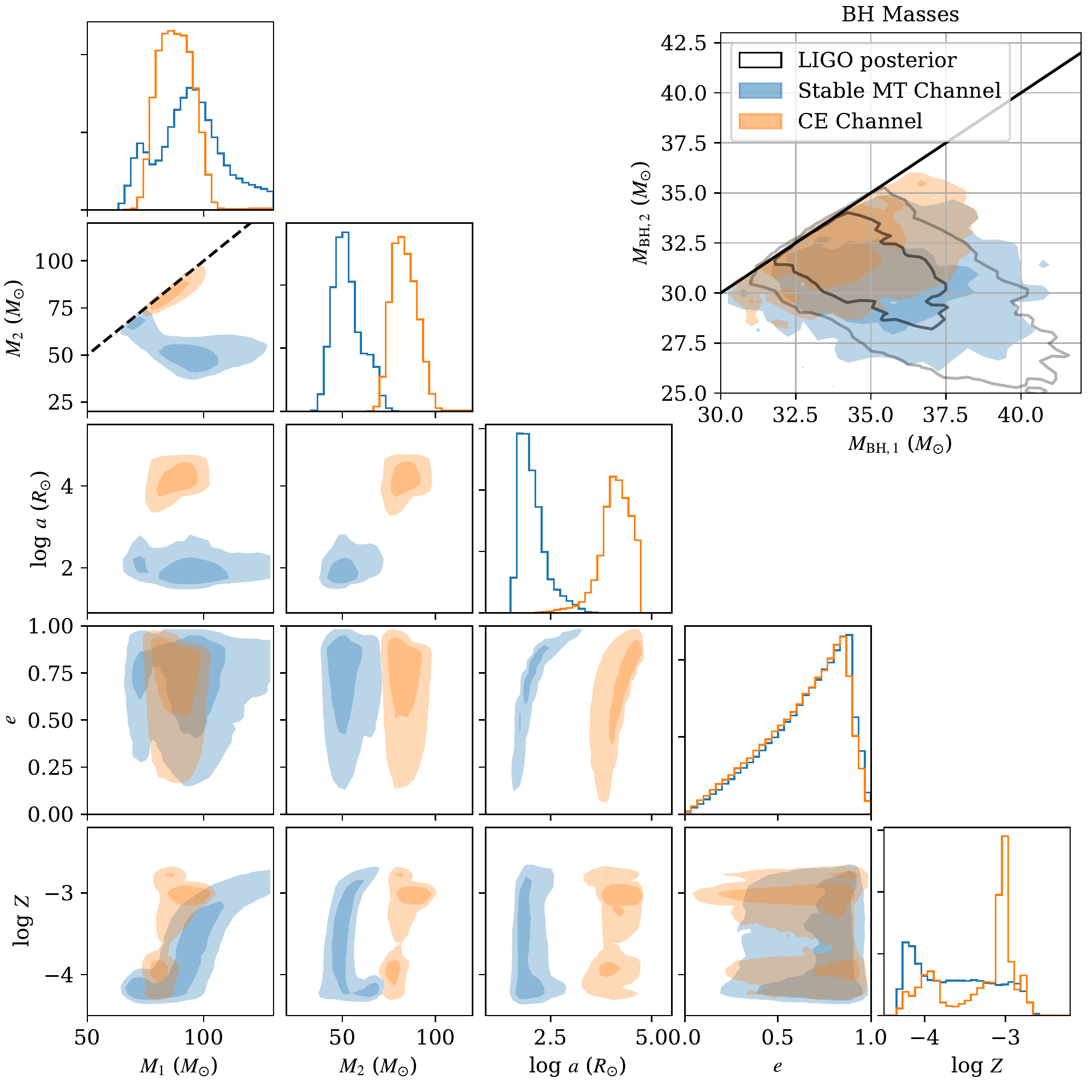}
\caption{ The ZAMS binary parameters forming GW150914 for our standard model. The top right panel compares the LSC posterior distribution (black contours) for the BH masses with our posterior distribution for the stable MT channel (blue contours) and the CE channel (orange contours) for $50\%$ and $90\%$ credible intervals. 
\label{fig:3}}
\end{center}
\end{figure*}

\section{Results}
\label{sec:results}

\subsection{Model Set-up}
\label{sec:model_setup}

To use \dart\ we must first initialize the samples in a high probability region of posterior space. Typically \dart\ does this by taking random draws in the high-dimensional parameter space, then initializing all the walkers in an $N$-ball around the point with the highest posterior probability. The case of GW150914 initialization is a challenge for two reasons: First, few binaries form binary BHs, and far fewer merge with conditions consistent with the observations of GW150914. Second, two separate formation channels could be responsible for forming GW150914, each with their own modes in posterior space which can cause difficulties for the proposal distributions of ensemble samplers like \emcee. 

We solve both problems using a two-step process for initialization. First, we run a \dart\ simulation with $320$ walkers for $20,000$ steps, selecting only for binaries that produce BH--BH systems; without any restriction for any characteristics related to GW150914 or even that the BH--BH systems ought to merge in a Hubble time, initialization using random draws is computationally reasonable. From the resulting sample of points in parameter space that produce BH--BH systems, we then calculate the likelihood function for each point for forming GW150914 from Eq.~\eqref{eq:likelihood_sum}. Most samples have GW150914 likelihoods that are effectively zero. However, a small subset of samples produce BH--BH systems consistent with the LVC constraints on GW150914's progenitor. From these samples we initialize a second set of separate \dart\ simulations, one for each channel, using the full likelihood in Eq.~\eqref{eq:likelihood_sum} to model GW150914. We find that there are two dominant formation channels forming GW150914's source, which we describe below. The presence of separate subchannels within these two formation channels causes some difficulty for the walkers, restricting their ability to efficiently move throughout the parameter space. We therefore use a parallel-tempering MCMC algorithm \citep{vousden2016} for each of these simulations, using $320$ walkers for $100,000$ steps. With our choice of a maximum temperature of $10$, we find the walkers are able to explore all the subchannels, largely without expanding to the alternative formation channel. In the cases where some walkers moved between both channels, we remove those that have transitioned in post-processing. From the posterior samples, we remove the first $10,000$ steps and derive statistical constraints on GW150914's formation from the remaining $90,000$ steps.

\subsection{Two Formation Channels}
\label{sec:formation_scenarios}

Regardless of our choice of model parameters, we always find two dominant formation channels: one in which MT always proceeds stably (stable MT channel) and one in which the system goes through at least one CE (CE channel). Figure~\ref{fig:2} provides a schematic for the evolution of a binary through each of these channels. After the initially more massive star in the binary has evolved off the main sequence, it overfills its Roche lobe leaving a helium star with a more massive main-sequence companion. Once the helium star collapses into a BH, the secondary then evolves off the main sequence and overfills its own Roche lobe. The bifurcation separating the two formation channels depends on whether this phase of MT is dynamically stable (stable MT channel) or unstable (CE channel). In both cases, the subsequent evolution proceeds similarly: the secondary follows the evolution of the primary, losing its entire hydrogen envelope and forms a helium star that eventually collapses to a BH; finally, gravitational radiation leads to the BHs merging. 

A variation on the CE channel depicted in Figure~\ref{fig:2} is realized in the \cosmic\ simulations, in which the two stars have similar enough masses that they evolve off the main sequence at similar times. When the (slightly) more massive star overfills its Roche lobe as a core-helium burning giant, the companion is also a giant star, leading to a double CE. Since it is not clear that such a case avoids a complete merger, and because the overall evolution is largely similar to the channel shown in the right-hand branch of Figure~\ref{fig:2}, throughout this work, we use the term CE channel to refer to both subchannels.

\begin{figure}
\begin{center}
\includegraphics[width=\columnwidth,angle=0]{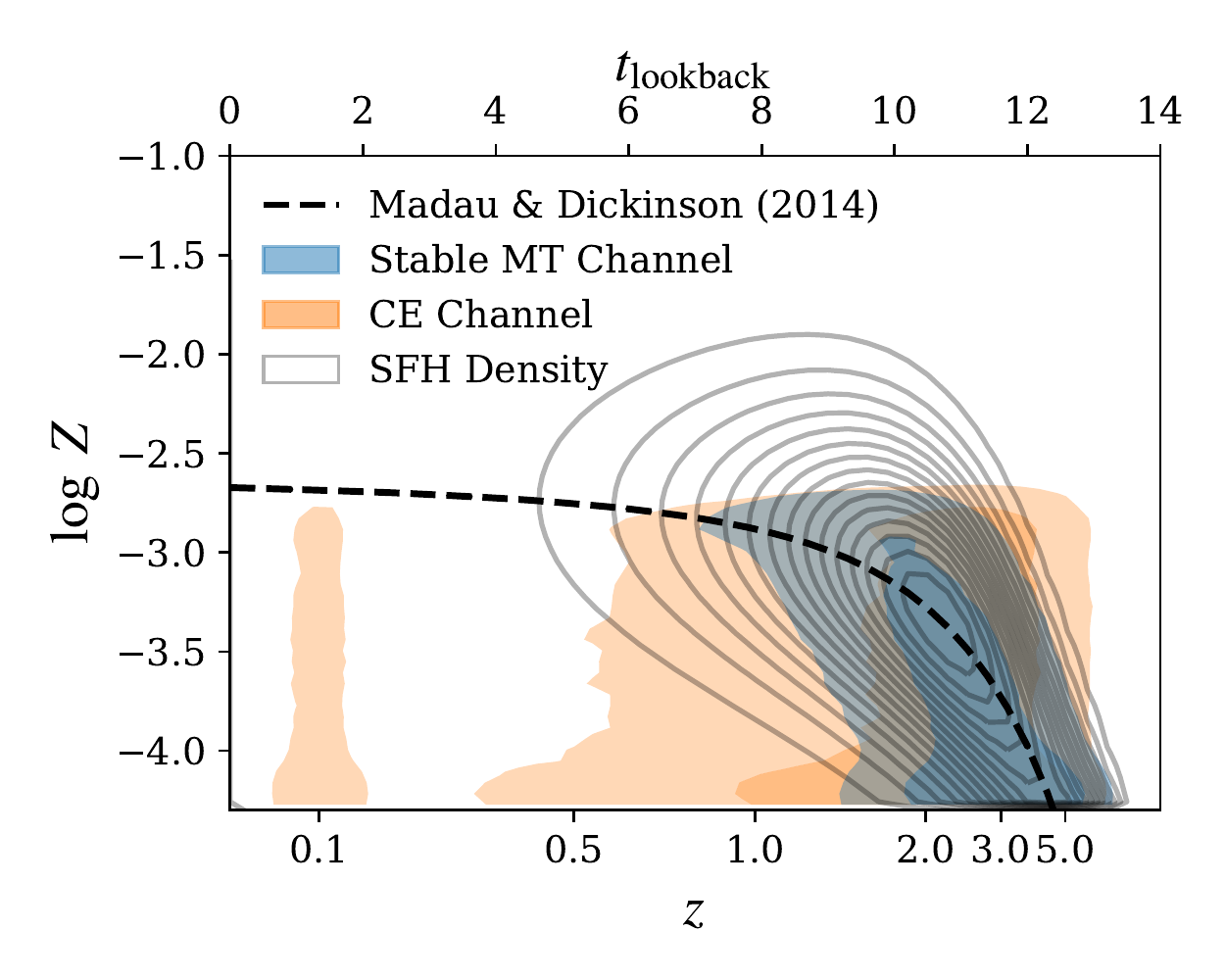}
\caption{ The relation between metallicity, lookback time and redshift from \citet{madau2014} is indicated by the dashed line. Gray contours indicate the star formation rate history in this space, and serve as our prior probability on $Z$ and $z$ (described in Section \ref{sec:cosmology}). We compare the prior probability against the posterior distributions from our two formation channels (Stable MT channel in blue; CE channel in orange) of the lookback time to the birth of the progenitor to GW150914. The BHs producing GW150914 are sufficiently massive that they require $\log Z \lesssim -2.5$. The metallicity-age relation from \citet{madau2014} strongly preferences such low metallicity systems to be born $\simeq$10--12 Gyr ago. \label{fig:4}}
\end{center}
\end{figure}

Despite the broad similarities between the stable MT and CE channels, the initial binary parameters are different. The corner plot in Figure~\ref{fig:3} compares the posterior samples between the two formation channels from our standard model for the ZAMS parameters: $M_1$, $M_2$, $\log a$, $e$, and $\log Z$. We ignore the eight model parameters associated with SN kicks as these provide little insight into formation; the posterior distributions for these parameters closely follow their priors, defined in Eq.~\eqref{eq:priors_2}. The difference between the two channels is most striking in the orbital separation of the progenitors. Since unstable MT shrinks the orbit by two orders of magnitude, whereas stable MT has a much smaller impact on the orbital separation, the GW150914-like binaries forming through the CE channel must start with an initially wider orbit; binaries going through this channel with smaller orbital periods at ZAMS will merge within the CE. The curved distributions in the log $a$--$e$ panel in Figure~\ref{fig:3} is a result of the fact that upon Roche lobe overflow {\tt cosmic} circularizes an initially eccentric binary so that angular momentum is conserved. There is therefore a degeneracy following $a(1-e^2)$ in the distributions in this panel, a result previously described by \citet{andrews2018}.

In addition to the orbital separation, the mass distribution between the two channels also differs. Shown clearest in the first panel of the second row, which compares the initial masses of the two stars, the CE channel typically forms from stars with similar initial masses of $\sim90\Msun$. As a comparison, the example binary progenitor to GW150914 described by \citet{belczynski2016} originated with two stars of masses $96.2 \Msun$ and $60.2 \Msun$ that evolved through the CE channel. The stable MT channel, on the other hand, forms from less massive stars, extending to mass ratios different from unity. The two separate maxima in the $M_1$ vs.\ $M_2$ panel of Figure~\ref{fig:3} are due to subchannels in the stable MT channel, depending on whether the stable MT phase is Case B or C.

While the initial eccentricity distribution is similar between the two formation channels, the metallicity distribution shows a stark difference. The stable MT channel shows a unimodal distribution, peaking at $Z=10^{-4}$ ($\simeq0.5\%~Z_{\odot}$), whereas the CE channel shows an additional peak at $Z=10^{-3}$ ($\simeq5\%~Z_{\odot}$). Since the metallicity of the Universe evolves over time, this double-peaked distribution in $\log Z$ corresponds to a bimodal distribution in the formation time of the system, a characteristic discussed by \citet{belczynski2016} in the context of the formation of GW150914 through the CE channel. 

To look more carefully at the distribution of $Z$, in Figure~\ref{fig:4} we show the covariance between log $Z$, $z$, and the birth time of the system, $t_{\rm lookback}$. Since metal-rich stars have stronger stellar winds and produce less massive BHs, the progenitor of GW150914 had to have been born with low metallicity \citep{GW150914_astrophysical_implications,belczynski2016}. Our results in the bottom right panel of Figure~\ref{fig:3} and in Figure~\ref{fig:4} show that, using our models \citep[based on SSE fits to stellar models;][]{hurley2000}, the progenitor to GW150914 must have been born with log $Z\lesssim-2.5$. Because we use the \citet{madau2014} relation between metallicity and lookback time, this low-metallicity requirement essentially ensures that the progenitor to GW150914 was born $\simeq$10--12~Gyr ago (although Figure \ref{fig:4} shows a small, low-probability region through the CE formation channel at an age of $\simeq1.5$~Gyr). 
The most probable formation time is consistent with \citet{lamberts2016}, who used a galactic redshift dependent mass--metallicity
relation \citep{mannucci2009}, but we find a different distribution of formation times.

Finally, we can use the posterior samples from our \dart\ simulations to derive updated constraints on the BH masses of GW150914's source. We show these constraints in the top right panel of Figure~\ref{fig:3} for both formation channels. As a consistency check, we overplot the LVC posterior distribution as grey contours. Neither channel adds significant improvements to the BH mass constraints, although compared with the stable MT channel, the CE channel forms BHs with mass ratios closer to unity.

\subsection{Branching Ratios}
\label{sec:branching_ratios}

\begin{figure}[ht]
\begin{center}
\includegraphics[width=\columnwidth,angle=0]{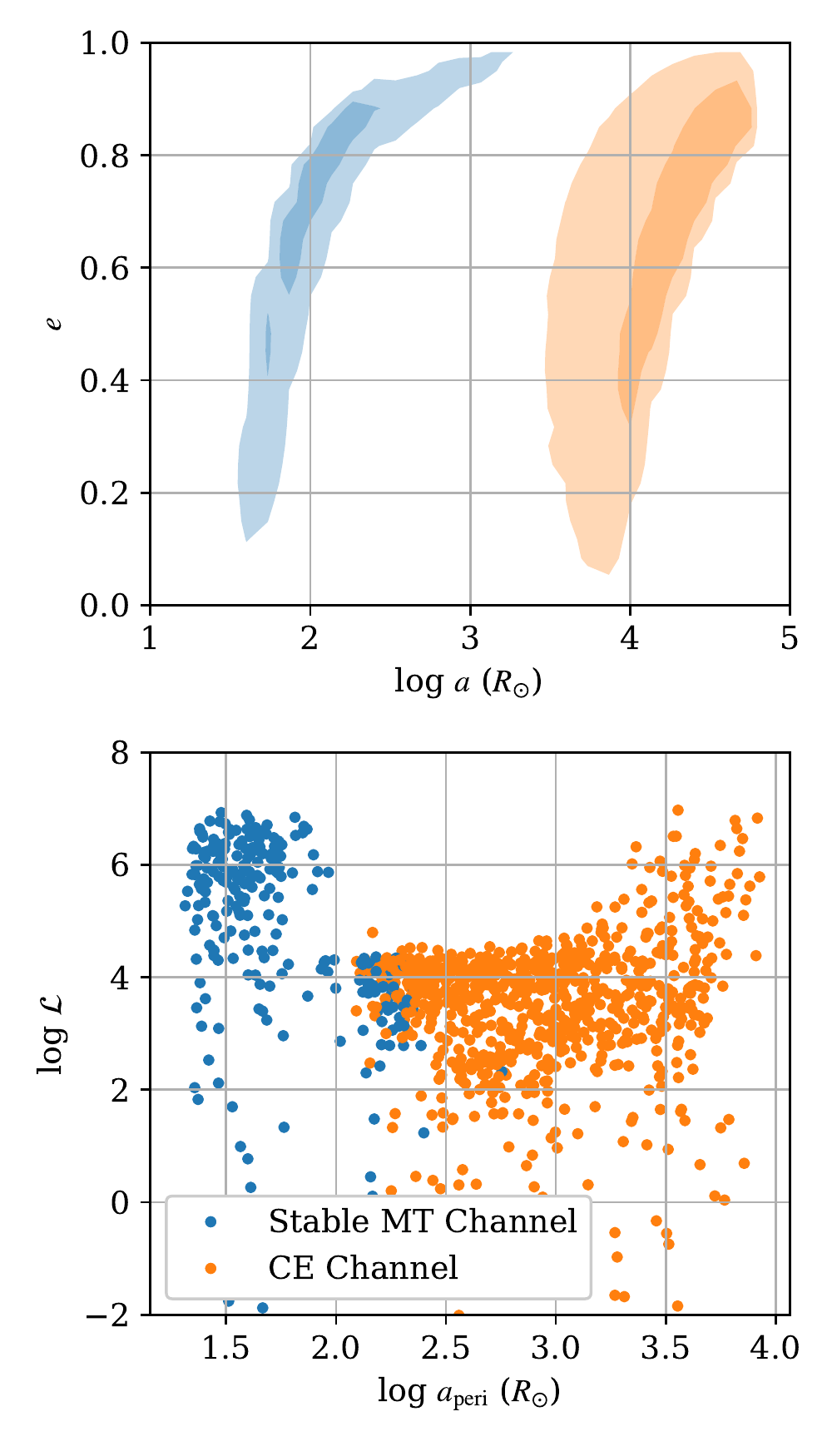}
\caption{ Orbital separation and eccentricity distribution at ZAMS for our standard model (top panel) for the stable MT channel (blue) and CE channel (orange). To determine the relative ratios between the two formation channels, we calculate the likelihoods as a function of the ZAMS periastron separation for a set of posterior samples from a \dart\ simulation that models the formation of binary BHs (bottom panel).  \label{fig:5}}
\end{center}
\end{figure}

While Figure~\ref{fig:2} shows the distribution of binaries formed through both principle formation channels, one cannot immediately glean the branching ratio between them. To do this, we follow the procedure outlined in Appendix~E of \citet{andrews2018}. We use the samples from the first step of our binary initialization procedure, our \dart\ model that produces only BH--BH binaries without any constraint from an LVC detection. These samples form a set of prior samples for any specific LVC observation. We then calculate the GW150914 likelihood, defined in Eq.~\eqref{eq:likelihood_sum} to each of these samples. In Figure~\ref{fig:5} we show the results of this calculation: the top panel shows the posterior distribution in $\log a$--$e$ from our GW150914 simulation (repeated from Figure~\ref{fig:3}), while the bottom panel shows the log likelihood as a function of the periastron separation at ZAMS for our prior sample of BH--BH binaries. The marker colors indicate each synthetic binary's formation channel. 

From Appendix~E of \citet{andrews2018}, the branching ratio between the two channels (indicated with the $A$ and $B$ subscripts) can be calculated from the sum of the likelihoods over the prior samples for both channels:
\begin{equation}
    \frac{f_A}{f_B} = \frac{\sum_1^{N_A} P(D | x_\mathrm{i}, \Lambda, C_A) }{\sum_1^{N_B} P(D | x_\mathrm{i}, \Lambda, C_B)},
\end{equation}
where $N_A$ and $N_B$ correspond to the number of prior samples going through each channel. The $C_A$ and $C_B$ in the likelihoods clarifies which channel the system had to have formed through. Normalizing the ratio between the two channels so they add to unity provides the branching probability for each channel $f_{\rm channel}$.

The bottom panel of Figure~\ref{fig:5} shows that there are two peaks in the likelihood function, one with $\log (a_{\rm peri}/R_\odot)\simeq1.5$ and one with $\log (a_{\rm peri}/R_\odot)\simeq3.5$. These correspond to the stable MT (blue markers) and CE (orange markers) channels, respectively. There are also a large number of points as part of the CE channel with intermediate orbital separations, with $\log (a_{\rm peri}/R_\odot)$ ranging from $2$--$3$. However, these points have likelihoods that are approximately three orders of magnitude lower than either of the two peaks at $\log (a_{\rm peri}/R_\odot)$ of $\simeq 1.5$ or $\simeq 3.5$. Therefore, these points do not contribute significantly to the posterior distribution in the top panel of Figure~\ref{fig:5}, leaving a large gap between the CE and stable MT contours.

\begin{table*}
  \centering
  \caption{ Posterior constraints on the prevalence and characteristics of the two dominant formation channels responsible for GW150914. Uncertainties define $90\%$ credible intervals. }
  \label{tab:1}
  \begin{tabular}{lc ccccc cc}
  \hline
  \multicolumn{2}{c}{Model $\Lambda$} & $f_{\rm channel}$ & $M_1$ & $M_2$ & log $a$ & log $Z$ & $M_{\rm BH, 1}$ & $M_{\rm BH, 2}$ \\
  &  &  & ($M_{\odot}$) & ($M_{\odot}$) & ($R_{\odot}$) &  & ($M_{\odot}$) & ($M_{\odot}$) \\
  \hline
  \multicolumn{9}{c}{Stable Mass Transfer Channel} \\
  \hline
  1 (std) & -- & 0.77 & 96$^{+37}_{-24}$ & 52$^{+16}_{-10}$ & 1.99$^{+0.73}_{-0.31}$ & $-3.63^{+0.81}_{-0.63}$ & 35.0$^{+5.3}_{-5.3}$ & 30.9$^{+6.1}_{-3.8}$ \\
  2 & $\alpha_{\rm CE}=0.5$ & 0.78 & 95$^{+36}_{-24}$ & 52$^{+16}_{-10}$ & 1.95$^{+0.78}_{-0.28}$ & $-3.67^{+0.84}_{-0.59}$ & 34.8$^{+5.5}_{-5.4}$ & 30.9$^{+6.4}_{-3.8}$ \\
  3 & $\alpha_{\rm CE}=5$ & 0.48 & 96$^{+36}_{-24}$ & 51$^{+15}_{-10}$ & 1.94$^{+0.89}_{-0.28}$ & $-3.59^{+0.78}_{-0.66}$ & 34.9$^{+5.2}_{-5.0}$ & 30.8$^{+5.7}_{-3.8}$ \\
  4 & Pessimistic CE & 0.84 & 95$^{+37}_{-24}$ & 52$^{+16}_{-10}$ & 2.01$^{+0.77}_{-0.33}$ & $-3.67^{+0.84}_{-0.59}$ & 34.9$^{+5.4}_{-5.5}$ & 30.9$^{+6.5}_{-3.9}$ \\
  5 & $f_{\rm acc}=1$ & 0.90 & 83$^{+23}_{-11}$ & 43$^{+10}_{-12}$ & 2.31$^{+1.11}_{-0.41}$ & $-3.41^{+0.59}_{-0.74}$ & 32.3$^{+4.6}_{-3.5}$ & 34.0$^{+4.0}_{-4.3}$ \\
  6 & $q_{\rm crit}=2$ & 0.02 & 91$^{+9}_{-8}$ & 47$^{+4}_{-4}$ & 1.87$^{+0.44}_{-0.24}$ & $-4.15^{+0.49}_{-0.13}$ & 37.4$^{+2.7}_{-2.9}$ & 30.2$^{+2.2}_{-2.6}$ \\
  7 & $\sigma_k=150$ km s$^{-1}$ & 0.94 & 89$^{+35}_{-15}$ & 61$^{+17}_{-18}$ & 2.65$^{+1.34}_{-0.75}$ & $-3.39^{+0.61}_{-0.73}$ & 34.1$^{+6.8}_{-4.6}$ & 32.1$^{+5.1}_{-6.1}$ \\
  \hline
  \multicolumn{9}{c}{Common Envelope Channel} \\
  \hline
  1 (std) & -- & 0.23 & 87$^{+13}_{-11}$ & 82$^{+13}_{-9}$ & 4.12$^{+0.51}_{-0.70}$ & $-3.12^{+0.33}_{-1.10}$ & 34.2$^{+3.7}_{-4.5}$ & 32.4$^{+2.9}_{-2.7}$ \\
  2 & $\alpha_{\rm CE}=0.5$ & 0.22 & 109$^{+35}_{-30}$ & 99$^{+36}_{-24}$ & 4.38$^{+0.32}_{-0.54}$ & $-2.77^{+0.13}_{-1.29}$ & 34.0$^{+3.7}_{-3.0}$ & 32.3$^{+2.9}_{-3.7}$ \\
  3 & $\alpha_{\rm CE}=5$ & 0.52 & 81$^{+10}_{-10}$ & 78$^{+10}_{-9}$ & 3.16$^{+1.06}_{-0.60}$ & $-3.12^{+0.32}_{-0.70}$ & 30.0$^{+3.8}_{-2.8}$ & 35.4$^{+3.3}_{-3.6}$ \\
  4 & Pessimistic CE & 0.16 & 89$^{+12}_{-12}$ & 84$^{+12}_{-10}$ & 4.16$^{+0.48}_{-0.53}$ & $-3.11^{+0.16}_{-1.05}$ & 34.5$^{+3.5}_{-3.1}$ & 32.1$^{+2.8}_{-2.6}$ \\
  5 & $f_{\rm acc}=1$ & 0.10 & 84$^{+13}_{-11}$ & 79$^{+14}_{-18}$ & 4.08$^{+0.55}_{-0.70}$ & $-3.08^{+0.33}_{-1.08}$ & 33.5$^{+4.0}_{-5.4}$ & 32.8$^{+4.1}_{-2.9}$ \\
  6 & $q_{\rm crit}=2$ & 0.98 & 107$^{+38}_{-25}$ & 82$^{+40}_{-20}$ & 4.09$^{+0.54}_{-0.64}$ & $-2.74^{+0.15}_{-0.38}$ & 35.1$^{+4.1}_{-3.7}$ & 31.0$^{+3.2}_{-4.5}$ \\
  7 & $\sigma_k=150$ km s$^{-1}$ & 0.06 & 87$^{+30}_{-10}$ & 82$^{+24}_{-9}$ & 4.12$^{+0.52}_{-0.59}$ & $-3.39^{+0.65}_{-0.73}$ & 34.3$^{+3.6}_{-3.1}$ & 32.0$^{+3.0}_{-2.9}$ \\
  \hline
  \end{tabular}
\end{table*}

\subsection{Model Variations}
\label{sec:model_variations}

In addition to our standard model, we run six other models varying several parameters which we expect to have significant impact on BH--BH formation. Differences between the models arise in the exact binary parameters forming through each channel as well as the branching ratios describing the number of systems evolving through each channel. We provide the posterior constraints on the ZAMS binary parameters for each model, separately for each channel, in Table~\ref{tab:1}. Listed errors define $90\%$ credible intervals. We discuss each of these models in turn below. The other parameters are fixed to their values in the standard model.

In our standard model, we set the CE efficiency to unity. However, based on recent CE hydrodynamic simulations suggesting that $\alpha_{\rm CE}$ may be as low as $0.5$ \citep[or possibly even lower;][]{law_smith2020} or as high as $5$ \citep{fragos2019}. Since the CE efficiency can have a significant effect on the merging BH--BH population \citep{dominik2012,barrett2018,giacobbo2018a,zevin2020b}, we test these two additional values. The BH--BHs evolving through the stable MT channel avoid CEs; therefore, differences in the ZAMS binary parameters in this channel when compared to our standard model are statistical. As expected, the CE channel shows some differences. Since a higher $\alpha_{\rm CE}$ corresponds to a more efficient envelope ejection mechanism, and therefore less orbital shrinkage, our $\alpha_{\rm CE}=5$ model has an initial orbital separation smaller by an order of magnitude. Additionally, this model has a branching ratio for the CE channel of $\simeq 52\%$, more that double that of our standard model. The $\alpha_{\rm CE}=0.5$ model, on the other hand, has characteristics similar to our standard model. 

Table~\ref{tab:1} shows an additional CE model (Pessimistic CE) in which we force all Hertzsprung gap stars to merge within a CE. The parameters describing the formation of these systems are nearly identical to our standard model, except fewer ($\simeq16\%$) systems form through the CE channel in this model. This is because in our fiducial model half of all binaries forming through this channel enter unstable MT on the Hertzsprung gap; the other half enter on the helium main sequence.

We run two additional models varying MT parameters. In our standard model, we assume half of all the mass lost by the donor during stable MT is incorporated by the accretor. This is the fraction assumed by \citet{belczynski2016} in explaining the formation of GW150914, and recently \citet{bouffanais2020} argue that the LVC data set implies this fraction $f_{\rm acc}$ cannot be less than $0.3$. Therefore, we additionally test a conservative MT model, in which no mass is lost from the system during MT ($f_{\rm acc}=1$). Compared with our standard model, Table~\ref{tab:1} shows that the initial secondary mass for both channels in this model is $\simeq5$--$10 \Msun$ less. Other model parameters show only minor differences. In our second MT model ($q_{\rm crit} = 2$) we set the critical mass ratio (donor mass divided by accretor mass) leading to a CE to be $2$ for all donor star types. Since this critical ratio is significantly less than our default value of $3$, many more binaries enter into a CE. As a result, this model shows the starkest contrast with all others. Table~\ref{tab:1} shows that $98\%$ of all viable samples form through a CE; systems that would normally form through the stable MT channel now enter unstable MT and are more likely to merge given their smaller orbital separations. This model is also characterized by higher initial primary masses, lower initial secondary masses, and BH mass ratios significantly different from unity.

As a final model, we apply kick velocities to BHs at birth using a Maxwellian prior with a dispersion velocity of $150~\mathrm{km\,s^{-1}}$ (compared with our standard model with $\sigma_k = 10~\mathrm{km\,s^{-1}}$), again these kick velocities are not moderated by supernova fallback. The increased kick velocity has the effect of broadening the distribution of possible initial orbital separations, as SN kicks can either expand or shrink the post-SN orbits, depending on the SN kick direction \citep{kalogera1996, andrews2019}. This model also shows a slight difference in the branching ratios, with nearly half of all binaries being formed through the CE channel.

\section{Discussion and Conclusions}
\label{sec:conclusion}

We have shown how statistically rigorous constraints can be constructed on individual gravitational-wave source progenitors. This approach offers a significant improvement over previous methods relying on some form of approximate Bayesian computation, as it takes into account the measurement uncertainties, including covariances between parameters. Using this method, we calculate the branching ratios for different evolutionary channels being responsible for the first LVC detection, GW150914, as well as the statistical constraints on the initial binary parameters responsible for formation through each channel. 

Under the assumption of formation through isolated binary evolution, the branching ratios from our standard model suggest that there is a $\simeq77\%$ chance GW150914 was formed through a stable MT channel and a corresponding $\simeq23\%$ chance that some form of a CE channel was involved. In the stable MT channel, the progenitor to GW150914 started as a low metallicity ($Z\simeq10^{-4}$) binary system with initial masses of $\simeq96 \Msun$ and $\simeq52 \Msun$ in an orbit of $\simeq 100 \Rsun$. In the CE channel, on the other hand, the binary was born with nearly equal mass stars ($\simeq 87 \Msun$ and $\simeq82 \Msun$) in a much wider orbit with a separation of $\simeq 10^4 \Rsun$ and low metallicity (bimodal distribution with peaks at $Z\simeq10^{-3}$ and $Z\simeq10^{-4}$). These masses are somewhat different from the example binary ($96.2 \Msun$ and $60.2 \Msun$) proposed by \citet{belczynski2016} to explain GW150914, although they are similar to the results from \citet{stevenson2017}, who find a total ZAMS mass $\gtrsim160\Msun$.

In all but one of our seven models the stable MT channel for the formation of GW150914 is preferred (in the $\alpha_{\rm CE}=5$ model the two channels are nearly equal). The one exception, our $q_{\rm crit}=2$ model (compared to our standard model, in which $q_{\rm crit}=3$ for most donor stars), finds the reverse, a $98\%$ chance that GW150914 was formed through a CE. Although this is the formation channel for GW150914 found as most probable by \citet{belczynski2016} and \citet{stevenson2017}, \citet{van_den_heuvel2017} argue that the most important MT phase---between a BH and a main-sequence donor---ought to be stable in many cases; these authors suggest a combination of the high accretor mass and strongly radiative envelope characterizing the donor stars implies that MT ought to be stable up to a mass ratio of $\simeq3.5$, even higher than our default model. A similar critical mass ratio is found by \citet{misra2020} for the production of ultra-luminous X-ray binaries and by Gallegos-Garcia et al.\ (in prep) for the production of BH--BH binaries, using detailed binary evolution simulations. We therefore conclude that GW150914's progenitor most probably avoided any CE phase.

Our method additionally allows for improved constraints on the BH masses responsible for the formation of GW150814 under the assumption that our model is an accurate representation of binary evolution; while the LVC posterior samples include an uninformative prior, our posterior samples account for a physically motivated prior based on binary evolution calculations. Our posterior constraints on the BH masses are largely consistent with the LVC constraints, a result of GW150914's high signal-to-noise ratio which implies that the posterior constraints are dominated by the likelihood.  Nevertheless, we find slight differences, depending on the formation channel responsible for GW150914; the CE channel predicts BHs with mass ratios close to unity, while the stable MT channel predicts GW150914 was formed from the merger of BHs with slightly different masses.

Our results here represent only a demonstration for a prototypical gravitational-wave event. With the latest release of the O3a data set \citep{GWTC-2}, there are now more than $40$ binary BH events, each of which can be analyzed using the same procedure. We expect that lower signal-to-noise ratio detections will have inferred parameters which are more sensitive to the choice of prior, and so may show the more significant updates when analysed with \dart. By analysing the entire population with \dart, we can begin to reconstruct the distribution of ZAMS properties of BH--BH progenitors and verify that these match expectations. We will pursue these in a forthcoming study.

Our results rely upon modelling the evolution of binary stars, which includes multiple uncertainties. We explored a range of these, quantifying how sensitive our inferences were to these assumptions. However, we can also calculate how well each of the models is supported by the observations, and hence infer the most probable parameter values for uncertain properties like the CE efficiency \citep{bavera2020a,wong2020,zevin2020b} or the fraction of mass accepted by the accretor \citep{bouffanais2020}. This requires an evidence integral calculation, which can be performed using a method similar to how we calculate the branching ratios in Section~\ref{sec:branching_ratios}. These constraints will become more precise when combining multiple observations together. The analysis could be further enhanced by incorporating information on the BH spins, which have been shown to be tracers of the formation history \citep{gerosa2013,rodriguez2016,bavera2020a}. For instance, in gravitational-wave measurements the spins are correlated with the mass ratio \citep{poisson1995,baird2013}, meaning that an astrophysical prior on the spins could impact the inferred masses and other progenitor properties. Additionally, the BH--BH merger rate, inferred by the LVC to be $23.9^{+14.9}_{-8.6}~\mathrm{Gpc^{-3}\, yr^{-1}}$ \citep{GWTC-2_population}, contains additional information on the formation physics \citep{barrett2018}. We will investigate the potential for detailed \dart\ model comparison in future work.

Our analysis currently assumes that any individual system formed through isolated binary evolution, rather than a dynamical, hierarchical triple, or more exotic formation channel. We have also not systematically explored isolated binary formation channels where two massive stars form chemically homogeneously \citep[][]{mandel2016, de_mink2016, eldridge2016, riley2020}. These channels have also been shown to be able to explain the formation of GW150914. Some form of hierarchical analysis \citep{santoliquido2020a,wong2020,zevin2020b} is required to determine the relative likelihoods between these different broadly defined formation channels. The \dart\ framework could be expanding to work with simulations of other channels to provide detailed inferences in these cases.

\acknowledgments

The authors thank Michael Zevin for a careful reading of the manuscript, and the anonymous referee for their comments.
We additionally thank the LVC for making their parameter-estimation posterior samples available through the Gravitational-Wave Open Science Center \citep{gwosc}.
JJA is funded by CIERA through a postdoctoral fellowship.
VK is supported by a CIFAR G+EU Senior Fellowship and Northwestern University. 
CPLB is supported by the CIERA Board of Visitors Research Professorship and NSF grant PHY-1912648.
AZ ackowledges funding from the European Research Council under the European Union’s Seventh Framework Programme (FP/2007-2013)/ERC Grant Agreement n. 617001 (A-BINGOS). This project has received funding from the European Union’s Horizon 2020 research and innovation programme under the Marie Sklodowska-Curie RISE action, grant agreement No 873089 (ASTROSTAT-II).
This work used computing resources at CIERA funded by NSF grant No.\ PHY-1726951, and resources and staff provided for the Quest high performance computing facility at Northwestern University, which is jointly supported by the Office of the Provost, the Office for Research, and Northwestern University Information Technology. 
This document has been assigned LIGO document number \href{https://dcc.ligo.org/LIGO-P2000503/public}{P2000503}.

\software{{\tt astropy} \citep{astropy}, {\tt emcee} \citep{emcee}, {\tt ptemcee} \citep{vousden2016}, {\tt kalepy} \citep{kelley2021}, {\tt NumPy} \citep{numpy}, {\tt SciPy} \citep{virtanen2020}, {\tt matplotlib} \citep{matplotlib}}

\bibliographystyle{aasjournal}
\bibliography{bibliography}

\end{document}